%% file: main.tex
\documentclass[conference]{IEEEtran}
\IEEEoverridecommandlockouts
\input{Commands}

\usepackage[binary-units,detect-all=true]{siunitx}
\DeclareSIUnit\decibelm{dBm}
\DeclareSIUnit\packets{packets}
\DeclareSIUnit\pct{percentile}
\AtBeginDocument{
	\DeclareSIUnit\bit{b}
	\DeclareSIUnit\bitpersec{bps}

	\newcolumntype{C}[1]{>{\centering\let\newline\\\arraybackslash\hspace{0pt}}m{#1}}
	\newcolumntype{L}[1]{>{\raggedright\let\newline\\\arraybackslash\hspace{0pt}}m{#1}}
	
}
\DeclareSIUnit[per-mode=symbol, per-symbol=p]\bps{\bit \per \second}
\DeclareSIUnit[per-mode=symbol, per-symbol=p]\kbps{\kilo \bit \per \second}\DeclareSIUnit[per-mode=symbol, per-symbol=p]\Mbps{\mega \bit \per \second}
\DeclareSIUnit[per-mode=symbol]\bpsHz{\bitpersec \per \hertz}

\usepackage[left=1.62cm,right=1.62cm,bottom=2.58cm,top=1.779cm]{geometry}

\def\BibTeX{{\rm B\kern-.05em{\sc i\kern-.025em b}\kern-.08em
		T\kern-.1667em\lower.7ex\hbox{E}\kern-.125emX}}

\author[1]{Pouria Paymard}
\author[2]{Abolfazl Amiri}
\author[2]{Troels E. Kolding}
\author[1, 2]{Klaus I. Pedersen}
\affil[1]{Department of Electronic Systems, Aalborg University, Aalborg, Denmark}
\affil[2]{Nokia, Aalborg, Denmark}
\affil[ ]{E-mail: pouriap@es.aau.dk}

\begin{document}
	\bstctlcite{IEEEexample:BSTcontrol}
	\graphicspath{{./Figures/}}
	\title{Enhanced Link Adaptation for Extended Reality Code Block Group based HARQ Transmissions}
	\maketitle
	
	\begin{abstract}\label{P1:Abstract}
		Extended Reality (XR) is one of the most important media applications in 5\textsuperscript{th} Generation (5G) and 5G-Advanced. XR traffic is characterized by high data rates with bounded latency constraints, which is challenging for bandwidth-constrained wireless systems. In this paper, we propose two new low-complexity enhanced Outer Loop Link Adaptation (eOLLA) algorithms that significantly improve the downlink system capacity in terms of satisfied XR users. The algorithms exploit the Code Block Group (CBG)-based Hybrid Automatic Repeat reQuest (HARQ) multi-bit feedback for minimizing the radio resource utilization in retransmissions by controlling the first and second block error operation points.
		Evaluation by means of both analytical assessment and realistic system-level simulations verifies that the proposed eOLLA algorithms increase system capacity by up to \SI{67}{\percent} compared to known OLLA algorithms with traditional transport block based HARQ.
	\end{abstract}
	
	\section{Introduction}\label{P1:Sec1:Introduction}
	Recent developments in 5\textsuperscript{th} Generation (5G) New Radio (NR) enable wireless support for immersive use cases such as eXtended Reality (XR) which is an umbrella term for augmented, virtual, and mixed reality applications \cite{3gpp.26.928}.
	XR traffic is characterized by high data rates (e.g., \SI{45}{\Mbps}) with bounded latency constraints of \SIrange{10}{15}{\milli\second} for the radio access network part and reliability targets on the order of \SI{99}{\percent} \cite{3gpp.38.838}. 
	Such requirement makes the design of efficient Link Adaptation (LA) and scheduling of XR traffic a challenging task in bandwidth-constrained wireless networks \cite{Hu2020Cellular_Connected}.
	
	In the 3rd Generation Partnership Project (3GPP) Release 17, a study item on XR over NR was recently conducted that included detailed definition of modeling and simulation assumptions for evaluation of XR performance, including definition of several Key Performance Indicators (KPI) such as XR system capacity \cite{3gpp.38.838}. Currently, a new 3GPP release 18 work item to further develop and standardize XR enhancements is ongoing \cite{3gpp.23.700-60}.
	In addition to the industrial 3GPP XR studies, there are numerous XR studies in the academic literature. As a few examples, the authors in \cite{taleb2021extremely} and \cite{clemm2020toward} provided surveys of XR over wireless systems, as well as an overview of low-latency immersive services and related solutions. In \cite{MAC_VR}, a multiuser Medium Access Control (MAC) scheduling scheme for the downlink (DL) virtual reality (VR) traffic over 5G with Multiple-Input and Multiple-Output (MIMO) is proposed, with the objective to maximize the number of VR clients while guaranteeing their Quality-of-Experience (QoE) by applying a delay-based scheduling policy and proper LA algorithm.
	
	The high data rates of XR applications lead to large Transport Blocks (TBs) to be sent to users. In 5G, the transmission of each TB is protected by asynchronous Hybrid Automatic Repeat reQuest (HARQ)\cite{asynchronous_HARQ}. In order to avoid retransmitting the full TB in case of errors, 5G supports Code Block Group (CBG)-based HARQ retransmission as earlier studied in \cite{khosravirad2017flexible,Reduced_CBG_HARQ_for_coexistence_with_urllc_traffic, yeo2017partial}. 
	Here, the principle is that Code Blocks (CBs) with their own Cyclic Redundancy Check (CRC) are arranged into CBGs.
	For each CBG, the receiver provides a separate Acknowledgment (ACK) or Negative Acknowledgment (NACK), such that only failed CBGs are retransmitted.
	
	Conducting accurate LA to choose the most efficient Modulation and Code Scheme (MCS) for each TB transmission is of importance to optimize the XR performance. This is typically conducted based on Channel Quality Indication (CQI) from the User Equipment (UE) \cite{Guillermo_2020_Channel_Quality_Feedback}. However, as CQI is subject to measurement and quantization errors, as well as other imperfections, an Outer Loop LA (OLLA) is typically introduced to compensate for the mentioned inaccuracies and control the Transport Block Error Rate (TBER) target \cite{OLLA2015self}. 
	As of today, the OLLA is designed for basic cases with traditional HARQ operation with a single ACK/NACK feedback per TB transmission. As an example, the authors in \cite{sarret2015dynamicOLLA} propose a dynamic OLLA algorithm for 5G. This OLLA algorithm is based on the mean and the standard deviation of the post Signal-to-Interference-plus-Noise Ratio (SINR) observations at the receiver to determine the optimal TBER target. Another OLLA scheme is introduced in \cite{blanquez2016eolla}, which adaptively modifies its step size to offset received CQI values according to the reception conditions. 
	
	Despite the rich 5G LA (and OLLA) studies in the literature, an efficient LA for XR use cases with CBG-based transmissions has not yet been extensively studied. In this paper, we propose two new enhanced OLLA (eOLLA) schemes for XR use cases with CBG-based HARQ transmissions that utilize the richer HARQ feedback from CBG-based transmissions. The objective of the proposed eOLLA schemes is to lower radio resource utilization of XR traffic, which  results in support for a higher number of satisfied XR users. We first present a theoretical analysis of the CBG-based transmission to motivate its potential advantages for XR use cases and to determine the desired block error rate operating point. This is used to motivate the design of new eOLLA algorithms that are assessed in a dynamic system level setting with multiple users, multiple cells, dynamic XR traffic, and accurate modeling of the major performance-determining radio access network functionalities in line with \cite{3gpp.38.838}. The presented results confirm that there are promising benefits to gain in terms of higher XR system capacity by adopting CBG-based HARQ schemes with the proposed eOLLA schemes.
	
	\section{Setting the Scene}\label{P1:Sec2:Setting the Scene}
	We consider a Time Division Duplex (TDD) 5G network with dense urban deployment, composed of $B$ Base Stations (BSs), each with three sectors and 200 meter inter-site distance \cite{3gpp.38.838}. We follow the 5G NR modeling assumptions including the physical layer numerology with 100 MHz carrier bandwidth configuration with 272 Physical Resource Blocks (PRBs) each consisting of 12 sub-carriers with 30 kHz Sub-Carrier Spacing (SCS). We assume a slot-based scheduling policy, where each slot consists of 14 OFDM symbols \cite{3gpp.38.211}.
	
	In line with \cite{3gpp.38.838}, the XR DL traffic is modeled as a sequence of video frames arriving at BS according to the considered video frame generation rate $\lambda_f$ frames per seconds (fps) and a random delay jitter $J_f$, which follows a truncated Gaussian distribution $J_f\sim \mathcal{TN}(\mu_j,\sigma_j,a_j,b_j)$ with mean $\mu_j$, variance $\sigma_j^{2}$, and non-zero interval $[a_j,b_j], a_j \leq b_j$. 
	The average inter-arrival time of video frame is denoted by $T_f=\frac{1}{\lambda_f}$ ms. In addition, the size of each video frame is a random variable which follows a truncated Gaussian distribution $S_f\sim \mathcal{TN}(\mu_s,\sigma_s,a_s,b_s)$ with mean $\mu_s$, variance $\sigma_s^{2}$, and non-zero interval $[a_s,b_s], a_s \leq b_s$.
	A video frame is denoted as a packet throughout the paper. 
	According to \cite{3gpp.38.838}, an XR UE should successfully receive more than X\% of its own packets within the Packet Delay Budget (PDB) to be a \textit{satisfied UE}. The system capacity is defined as the maximum number of UEs per cell with at least Y\% of UEs marked as satisfied. 
	\section{An Analytical Assessment of CBG-based Transmissions}\label{P1:Sec3:Analytical_Assessment}
	\subsection{Basic principles}\label{P1:Sec3:SubsecS:Basic principle}
\begin{figure}[t]
		\centerline{\includegraphics[width=0.8\linewidth,trim={7.3cm 4cm 0cm 0cm},clip]{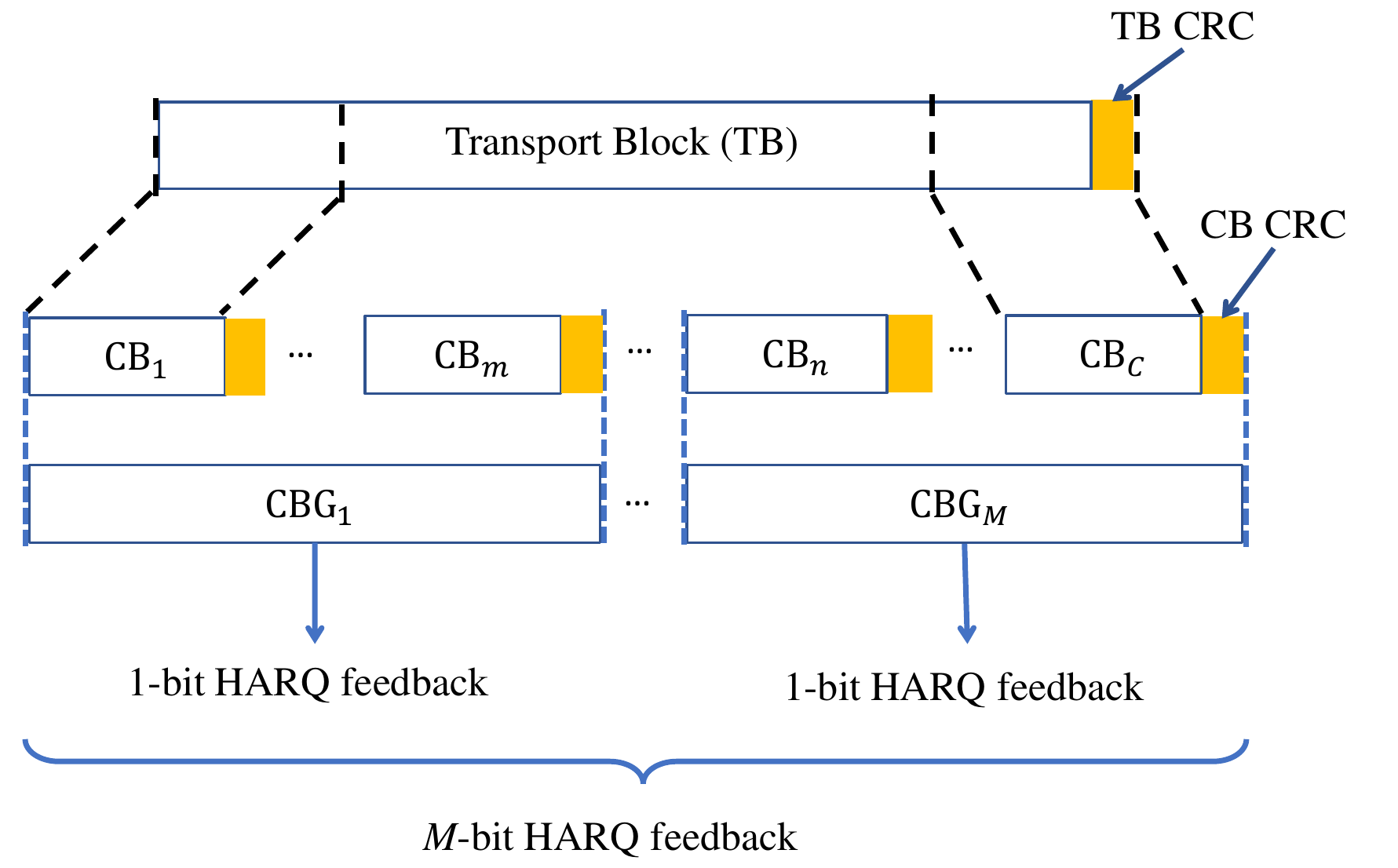}}
		\caption{The construction of $M$ CBGs per TB, each consisting of $C$ CBs.}
		\label{P1:Fig:TB_to_CBG}
	\end{figure}

	A TB is composed of one or several CBs of maximum size $8448$ bits that are separately encoded and have their own CRC \cite{3gpp.38.214}. Multiple CBs are grouped into a CBG. 
	A multi-bit HARQ feedback is provided in which each bit
	can indicate failure of CBG.
	In case of error in any CB of a certain CBG, a NACK is generated and the BS retransmits the failed CBG.
	Fig. \ref{P1:Fig:TB_to_CBG} illustrates the relationship between TB, CB, CBG, as well as multi-bit HARQ feedback corresponding to CBGs.
	According to \cite{3gpp.38.331}, the maximum number of CBGs per TB is configured as $N \in \left\lbrace 2, 4, 6, 8 \right\rbrace$ for Physical Downlink Shared Channel (PDSCH). 
	The number of CBGs in the TB is obtained by $M = \min(N,C)$, where $C$ is the number of CBs in the TB.
	
\begin{figure}[t]
		\centerline{\includegraphics[width=0.9\linewidth,trim={0.8cm 0.0cm 1.3cm 0.2cm},clip]{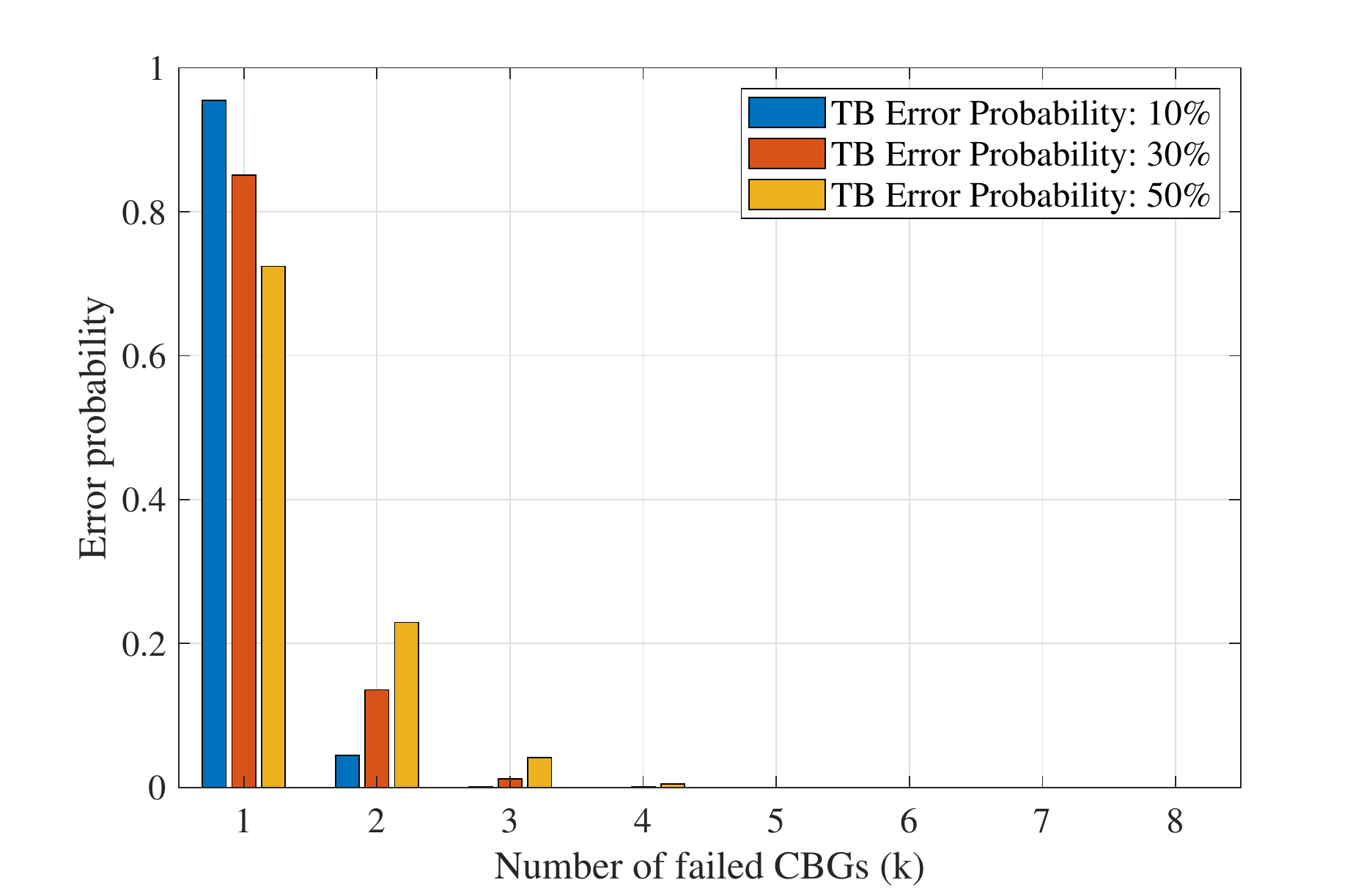}}
		\caption{The probability of number of erroneous CBGs for different TB error probabilities when there are $8$ CBGs in a TB.}
		\label{P1:Fig:failedCBGs_BLER}
	\end{figure}

	The error probability for a specific TB and CBG are denoted by $P_{e_\text{TB}}$ and $P_{e_\text{CBG}}$, respectively. 
	If independent and identically distributed (i.i.d.) CBG error probabilities are assumed within a TB, $P_{e_\text{TB}}$ is given by \cite{khosravirad2017flexible, Reduced_CBG_HARQ_for_coexistence_with_urllc_traffic, yeo2017partial}:
	\begin{equation}\label{P1:Eq:accurate_and_approx_TBLER}
		P_{e_\text{TB}}= 1- \prod_{i=1}^{M} \left(1-P^i_{e_\text{CBG}}\right) \approx 1-\left(1-P_{e_\text{CBG}}\right)^M,
	\end{equation}
	where $P^i_{e_\text{CBG}}$ is the error probability of $i^{\text{th}}$ CBG and the approximate equality holds if the CBGs roughly have equal sizes \cite{lagen2020new}.
	By doing simple mathematical manipulations, the CBG error probability is
	derived from \eqref{P1:Eq:accurate_and_approx_TBLER} as:
	\begin{equation}\label{P1:Eq:CBGER}
		P_{e_\text{CBG}}=1-\left(1-P_{e_\text{TB}}\right)^{\frac{1}{M}}.
	\end{equation}
	Let $n^{\text{CBG}}_e$ represent the number of failed CBGs in a TB. Hence, the probability of having $k$ failed CBGs, $P^{\text{Fail}}_{\text{CBG}}(n^{\text{CBG}}_e=k, \forall k \in\left\lbrace 1, 2, \dots, M \right\rbrace)$, can be expressed as
	\begin{equation}\label{P1:Eq:EP_num_failed_CB}
		P^{\text{Fail}}_{\text{CBG}}(n^{\text{CBG}}_e=k, P_{e_\text{CBG}})={M \choose k} \left(P_{e_\text{CBG}}\right)^k \left(1-P_{e_\text{CBG}}\right)^{M-k}.
	\end{equation}
	Using \eqref{P1:Eq:EP_num_failed_CB}, Fig. \ref{P1:Fig:failedCBGs_BLER} illustrates the probability of number of failed CBGs in a TB for different TB error probabilities.
	As an example, if a HARQ-NACK is received, the probability of having two or less failed CBG, is about \SI{99.8}{\percent} when $P_{e_\text{TB}}=\SI{10}{\percent}$.
	This confirms that the network can save radio resources by applying CBG-based HARQ retransmissions as is further studied in the following subsection.
	\subsection{CBG-based Radio Resource Efficiency for XR}\label{P1:Sec3:SubsecB:Gain Evaluation}
\begin{figure}[t]
		\centerline{\includegraphics[width=\linewidth,trim={0.5cm 0.0cm 0.9cm 0.3cm},clip]{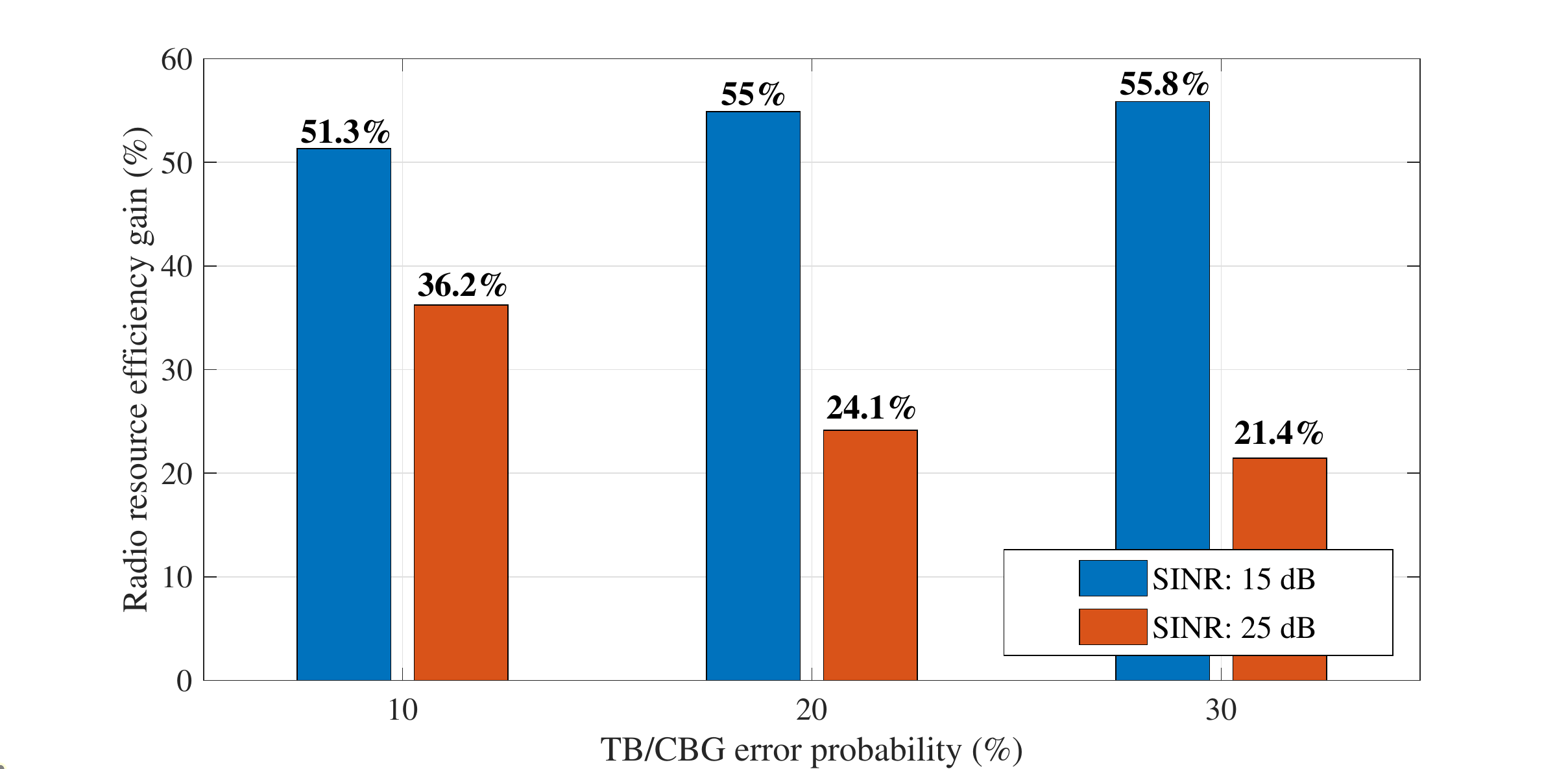}}
		\caption{The achievable RREG versus TB/CBG error probability for two different SINR regimes.}
		\label{P1:Fig:ResourceEfficiencyGain_TBERandSINR}
	\end{figure}

	The radio resource efficiency for transmission of a XR packet is a function of the first transmission block error probability (which is a result of the user experienced SINR and the selected modulation and coding scheme – MCS), and the number and size of potential retransmission. In the following we compare TB and CBG based transmissions to assess the radio resource efficiency gain potential of using optimized CBG-based XR transmissions. The average number of CBGs that are retransmitted after the first transmission (1st-TX) can be expressed as
	\begin{equation}\label{P1:Eq:RU_1stTx}
		R^{\text{1st}}_{\text{CBG}}(P_{e_\text{CBG}}) =	P_{e_\text{TB}}\left( \sum\limits_{n^{\text{CBG}}_e=1}^{M} n^{\text{CBG}}_e . P^{\text{Fail}}_{\text{CBG}}(n^{\text{CBG}}_e, P_{e_\text{CBG}})\right).
	\end{equation}
	While the average number of CBGs that are retransmitted after the second transmission (2nd-TX) is given by
	\begin{equation}\label{P1:Eq:RU_2stTx}
		R^{\text{2nd}}_{\text{CBG}}(P_{e_\text{CBG}})
		=\textit{f}\left( P_{e_\text{TB}} \right) \left(\sum\limits_{n^{\text{CBG}}_e=1}^{R^{\text{1st}}_{\text{CBG}}} n^{\text{CBG}}_e . P^{\text{Fail}}_{\text{CBG}}(n^{\text{CBG}}_e, P_{e_\text{CBG}})\right),
	\end{equation}
	where $\textit{f}\left( P_{e_\text{TB}} \right)$ is the error probability of 2nd-TX, which is a function of 1st-TX error probability $P_{e_\text{TB}}$.
	Given the use of HARQ soft combining, we assume that the 2nd transmissions are successful. We define the Radio Resource Efficiency Gain (RREG) as the percentage of radio resources which are saved by CBG-based transmissions in comparison with TB-based transmissions. Hence, the RREG, $\eta^{\text{RREG}}$, is a function of the respective number of CBG transmissions and the selected MCS index, which can be formulated as
	\begin{equation}\label{P1:Eq:RSR}
		\eta^{\text{RREG}} = \hspace*{-0.05cm} 100 \hspace*{-0.05cm}
		\left(1-\frac{(\xi^{\text{MCS}}_{\text{CBG}})^{-1} . (M+R^{\text{1st}}_{\text{CBG}} + R^{\text{2nd}}_{\text{CBG}}) }{(\xi^{\text{MCS}}_{\text{TB}})^{-1} . \left(M + P_{e_\text{TB}}. M + \textit{f}\left( P_{e_\text{TB}} \right) . M \right)}\right)\hspace*{-0.10cm},
	\end{equation}
	where
	$\xi^{\text{MCS}}_{\text{CBG}}$ and $\xi^{\text{MCS}}_{\text{TB}}$ are the resource efficiency achieved by the respective MCS index selection in CBG-based and TB-based transmissions, respectively. Notice that we assume that the MCS for any HARQ retransmission is the same as for the first transmission. Based on standard 5G NR link level performance results of the block error probability versus SINR for different MCS indices, we know the resource efficiency for a given MCS index, SINR, and the block error probability.
	In Fig. \ref{P1:Fig:ResourceEfficiencyGain_TBERandSINR}, we plot $\eta^{\text{RREG}}$ from \eqref{P1:Eq:RSR} versus the first transmission block error probability for two different UE receiver experienced SINR regimes of 15 dB and 25 dB. 
	In order to calculate the resource utilization in TB-based transmissions (i.e., denominator of \eqref{P1:Eq:RSR}), we set three different TB error probabilities, $P_{e_\text{TB}}\in\left\lbrace 10 , 20 ,30 \right\rbrace$, for the TB-based transmissions.
	In addition, we apply three different CBG error probabilities, $P_{e_\text{CBG}}\in\left\lbrace 10 , 20 ,30 \right\rbrace$, for the CBG-based transmissions to determine the resource utilization in TB-based transmissions (i.e., nominator of \eqref{P1:Eq:RSR}).
	It can be seen that a significant RREG is achieved by CBG-based transmissions. Inspecting both Fig. \ref{P1:Fig:failedCBGs_BLER} and Fig. \ref{P1:Fig:ResourceEfficiencyGain_TBERandSINR}, it is observed that to fully unleash the performance benefits of CBG-based transmission, the initial transmission block error probability should be higher as compared to the case with TB-based transmissions. This motivates our new eOLLA schemes, explained in the following section, where controlling the desired block error probability operation point is a key element to achieve resource-efficient transmissions. It is also observed from Fig. \ref{P1:Fig:ResourceEfficiencyGain_TBERandSINR} that the RREG is higher for the low SINR regime of 15 dB as compared to the high SINR of 25 dB. The explanation for this behavior is that the system operates in the linear part of the link adaptation dynamic range for the 15 dB SINR, while at 25 dB SINR use of the maximum modulation schemes (64QAM and 256QAM) is more pronounced and less opportunities for link adaptation adjustments.
	
	\section{CBG-based Enhanced OLLA Algorithms}\label{P1:Sec4:CBG-based Link Adaptation}
	Inspired by the results in the previous section, it is evident that appropriate LA is needed to gain from the CBG-based transmissions, where especially the error rate for the CBGs must be controlled, as compared to only controlling the error rate of first transmission TBs as is the case for traditional LA with standard OLLA. We therefore propose two eOLLA algorithms with such objectives in this section. We rely on the currently supported 5G CQI schemes that guide the BS to use a certain MCS index $m$ for its transmissions as follows, 
	\begin{equation}\label{P1:Eq:BLER_based_MCS_selection}
		\varphi^*=\arg\max\limits_{\varphi}\left\lbrace R_{\varphi} | \text{TBER} \leq \text{TBER}^{\text{Target}} \right\rbrace,
	\end{equation}
	where $R_{\varphi}$ is the supported data rate by using MCS index $\varphi$ while the TBER does not exceed the TBER target.
	In practice, the UE implements \eqref{P1:Eq:BLER_based_MCS_selection} by measuring its experienced post-detection SINR and using an internally stored table of its performance in terms of TBER versus SINR for each of the supported MCS indices that can be indicated in the CQI report. The CQI report is designed such that there is a constant equivalent SINR offset of X dBs between the MCS indices that can be indicated in the CQI report, where X typically is on the order of 1-2 dB, depending on the 5G NR configuration of the CQI report (see more in \cite{Guillermo_2020_Channel_Quality_Feedback}).  Due to various CQI imperfections such as UE measurement and quantization errors, reporting delays, and interference variations, the received CQI from the UE is offset by an OLLA offset, $\Delta_{\text{Offset}}$. The so-called effective  compensated SINR after applying the OLLA offset is therefore expressed as
	\begin{equation}\label{P1:Eq:OLLA_offset_CQI}
		\gamma_{\text{Eff}} [\text{dB}] = \gamma_{\text{CQI}} [\text{dB}] - \Delta_{\text{Offset}} [\text{dB}],
	\end{equation}
	where $\gamma_{\text{CQI}}$ is the reported CQI value to the BS. While the traditional OLLA algorithm for TB-based HARQ transmission only acts on a Boolean HARQ ACK/NACK feedback for the TB-based transmission to control the initial error rate of the full TB, we propose two new eOLLA algorithms with different objectives. In the first scheme, summarized in Algorithm \ref{P1:Alg:Traditional_OLLA_per_CBG}, we consider the case where the eOLLA aims at controlling the first transmission error rate of the individual CBGs. In the second scheme, summarized in Algorithm \ref{P1:Alg:eOLLA_ResidualBLER_2ndTx}, we aim to control the residual error rate of second transmissions, i.e., equivalent to first HARQ retransmissions. Notice that for both schemes, we take advantage of the multi-bit HARQ feedback (i.e., one ACK/NACK per CBG) to help improve the convergence and precision of the eOLLA schemes.
	\begin{algorithm}[t]
		\caption{Controlling the CBGER of 1st-TX}\label{P1:Alg:Traditional_OLLA_per_CBG}
		\begin{algorithmic}[1]
			\STATE
			\textbf{Initialize}\\
			Initial OLLA offset value \textrightarrow $\Delta_{\text{Offset}}$, \\ 
			Step up size \textrightarrow $\Delta_{\text{up}}$, \\
			Step down size \textrightarrow $\Delta_{\text{down}}$.
			\IF {a $M$-bit HARQ feedback includes $F\geq 0$ bits NACKs}
			\STATE
			$\Delta_{\text{Offset}}=\Delta_{\text{Offset}}-\left( \Delta_{\text{down}}\times\frac{M-F}{M}\right)  + \left( \Delta_{\text{up}}\times\frac{F}{M}\right)$.  \label{dynamic_step_up2}
			\ENDIF
			\STATE
			The effective SINR $\gamma_{\text{Eff}}$ is derived from \eqref{P1:Eq:OLLA_offset_CQI}. \\
			\STATE
			Based on $\gamma_{\text{Eff}}$, the proper MCS index is selected by \eqref{P1:Eq:BLER_based_MCS_selection}.
			\STATE 
			The CBGER will converge to \eqref{P1:Eq:CBGER_Convergence}.		
		\end{algorithmic}
	\end{algorithm}
	
	Algorithm \ref{P1:Alg:Traditional_OLLA_per_CBG} takes advantage of the CBG-based HARQ multi-feedback such that the increase/decrease of the $\Delta_{\text{Offset}}$ is scaled by the ratio of failed CBGs in the TB (see Step \ref{dynamic_step_up2}). This allows more accurate control and also offers faster convergence of the eOLLA as compared to the traditional OLLA scheme that always perform large increases of the $\Delta_{\text{Offset}}$ independent on whether the TB is subject to few or many failed CBGs. It works similarly as traditional OLLA, except that the eOLLA offset is cumulatively adjusted based on all the ACK/NACKs from the CBGs in first transmissions.
	The parameters of the eOLLA algorithms, i.e.,  $\Delta_{\text{up}}$, $\Delta_{\text{down}}$ and  the initial eOLLA offset, are set as network configuration parameters. It can be shown that the 1st-TX CBG Error Rate (CBGER), i.e., the number of failed CBGs divided by all CBGs in a certain TB, will converge to the following, and hence can be controlled via the setting of parameters $\Delta_{\text{up}}$ and $\Delta_{\text{down}}$,
	\begin{equation}\label{P1:Eq:CBGER_Convergence}
		\text{CBGER}_\text{1st-TX}^\text{Target}=\frac{1}{1+\frac{\Delta_{\text{up}}}{\Delta_{\text{down}}}}.
	\end{equation}
	
\begin{figure}[t]
		\centerline{\includegraphics[width=0.9\linewidth,trim={0cm 0cm 2cm 0cm},clip]{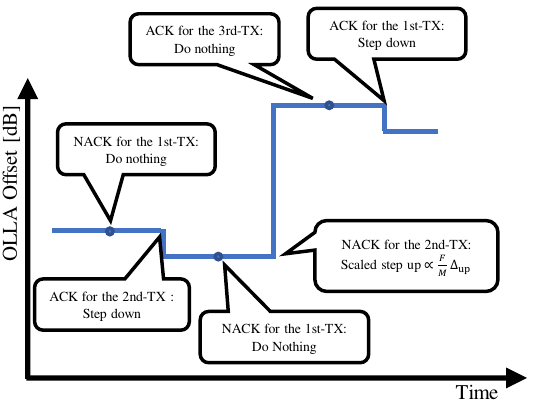}}
		\caption{An example of how the OLLA offset is adjusted over time for Algorithm \ref{P1:Alg:eOLLA_ResidualBLER_2ndTx}.}
		\label{P1:Fig:ResidualBLER_2ndTx}
	\end{figure}

	\begin{algorithm}[t]
		\caption{Controlling the Residual TBER of 2nd-TX}\label{P1:Alg:eOLLA_ResidualBLER_2ndTx}
		\begin{algorithmic}[1]
			\STATE
			\textbf{Initialize}\\
			Initial OLLA offset value \textrightarrow $\Delta_{\text{Offset}}$, \\ 
			Step up size \textrightarrow $\Delta_{\text{up}}$, \\
			Step down size \textrightarrow $\Delta_{\text{down}}$.
			\IF {ACK is received from a 1st-TX or 2nd-TX}
			\STATE
			$\Delta_{\text{Offset}}=\Delta_{\text{Offset}}-\Delta_{\text{down}}$.
			\ELSIF {$F$ HARQ NACKs are received in a 2nd-TX}
			\STATE 
			$\Delta_{\text{Offset}} = \Delta_{\text{Offset}} + \Delta_{\text{up}}\times\frac{F}{M}, F \geq 1$. \label{dynamic_step_up}
			\ENDIF
			\STATE
			The effective SINR $\gamma_{\text{Eff}}$ is derived from \eqref{P1:Eq:OLLA_offset_CQI}. \\
			\STATE
			Based on $\gamma_{\text{Eff}}$, the proper MCS index is selected by \eqref{P1:Eq:BLER_based_MCS_selection}.
			\STATE 
			The residual TBER of the 2nd-TX will converge to \eqref{P1:Eq:2ndTBER_Convergence}.
		\end{algorithmic}
	\end{algorithm}
	Algorithm \ref{P1:Alg:eOLLA_ResidualBLER_2ndTx} controls the residual TBER of the 2nd-TX as described by the example in Fig. \ref{P1:Fig:ResidualBLER_2ndTx}. This is of relevance to XR use cases, where it is important to control the residual TBER, while of less importance to accurately control the 1st-TX to equal a specific TBER target. This solution is realized by monitoring the HARQ multi-bit feedback from both the first and the second HARQ transmissions (i.e., corresponding to the first retransmission). Here, the eOLLA Offset is only modified if either the 1st-TX or 2nd-TX is successfully decoded or the 2nd-TX is not decoded. Otherwise, no actions are taken if the 1st-TX is in error. The $\Delta_{\text{Offset}}$ increases for the case of detected errors in 2nd-TXs and decreases upon receiving ACKs from either 1st or 2nd transmissions. The increase of step size is weighted proportional to the CBGs in error to reflect the cost of subsequent retransmission of those failed CBGs.
	With this approach, assuming operation within the link adaptation dynamic range, the residual TBER after the 2nd-TX will converge to
	\begin{equation}\label{P1:Eq:2ndTBER_Convergence}
		\text{TBER}_\text{2nd-TX}^\text{Target}=\frac{1}{1+\frac{\Delta_{\text{up}}}{2\times\Delta_{\text{down}}}}.
	\end{equation}
	
	\section{System-level Performance Evaluation}\label{P1:Sec5:Performance Evaluation}
	The performance of the proposed CBG-based eOLLA algorithms is evaluated by means of dynamic system-level simulations, where all the major performance-determining radio access network mechanisms are accurately accounted for. We follow the simulation guidelines for the dense urban deployment scenario provided in \cite{3gpp.38.838}. The baseline for assessing the performance benefits of the proposed CBG-based transmissions with the eOLLA algorithms is use of traditional TB-based traditional OLLA with a first transmission TBER target of either \SI{10}{\percent} or \SI{30}{\percent}. The system model outlined in Section \ref{P1:Sec2:Setting the Scene} is followed, adopting the simulation assumptions in Table \ref{P1:Table:System_Parameters}. 
	\begin{table}[t]
		\centering
		\caption{Summary of System-level Evaluation Parameters}
		\begin{tabular}{|c|c|}
			\hline
			\textbf{Parameter} & \textbf{Setting} \\ \hline
			\multicolumn{2}{|c|}{\textbf{General}} \\ \hline
			Simulation time & 10 seconds \\ \hline
			Simulation runs  & 15 runs per result \\ \hline
			Deployment layout & Dense Urban \\ \hline
			Number of cells & 21 (7 BSs with 3 sectors)  \\ \hline
			Inter-site Distance & 200 m \\ \hline
			Channel model & Urban Macro (UMa) \\ \hline
			User distribution & Even UEs per cell\\ \hline
			TDD Frame structure & DDDSU \\ \hline
			Scheduler & Proportional Fairness \\ \hline
			Carrier frequency & 4 GHz  \\ \hline
			System Bandwidth &  100 MHz \\ \hline
			SCS & 30 kHz \\ \hline
			BS height & 25 m  \\ \hline
			BS power & 50 dBm \\ \hline
			\multirow{2}*{BS antenna} & 1 panel with 32 elements \\
			& (8 × 2 and 2 polarization) \\ \hline
			UE speed & 3 km/h  \\ \hline
			PHY processing delay & 6 OFDM symbols \\ \hline
			CSI & Periodic CQI on 2 ms period  \\ \hline
			Rank & Adaptive (between 1 \& 2)  \\ \hline
			\multicolumn{2}{|c|}{\textbf{XR Traffic Model}} \\ \hline
			Frame rate & 60 fps \\ \hline
			Jitter distribution & $\mathcal{TN}(0, 2, -4, 4)$ ms  \\ \hline
			Frame size distribution & $\mathcal{TN}(62.5, 6.25, 31.25, 93.75)$ kByte \\ \hline
			Required data rate & 45 Mbps \\ \hline
			PDB & 10 ms \\ \hline
			Reliability & \SI{99}{\percent} \\ \hline
			\multicolumn{2}{|c|}{\textbf{Link Adaptation}} \\ \hline
			Maximum number of & \multirow{2}*{3} \\ 
			HARQ retransmissions &  \\ \hline
			HARQ scheme & Chase combining \\ \hline
			Highest Modulation & 256QAM \\ \hline
			Lowest Modulation & QPSK \\ \hline
			Initial Offset & 0 dB \\ \hline
			Maximum OLLA Offset & 15 dB \\ \hline
			Minimum OLLA Offset & -25 dB \\ \hline
		\end{tabular}
		\label{P1:Table:System_Parameters}
	\end{table}
	Simulations are conducted at symbol level (time-domain) and sub-carrier resolution (frequency domain), assuming 14 OFDM symbols per TTI, where the first symbol of each TTI carries control overhead. For each transmission, the effective SINR is computed and used to determine if the TB and CBGs are correctly received in line with models also used in \cite{Preemptive_CBlayouts}. This considers the use of MIMO transmissions, UE receivers with interference rejection combining, interference from neighboring cells, etc. Each cell only schedules transmissions on the PRBs for which it has data to transmit, assuming standard proportional fair scheduling. Periodic CQI reporting is assumed for LA purposes, and we assume error-free ACK/NACK feedback from the UEs. To determine the supported XR capacity, we run simulations at different load levels in terms of number of XR UEs per cell. To obtain statistical reliable performance results, we run multiple long simulations for each load level, so we have sufficient statistics from at least 1575 XR calls of $600$ XR packets to assess the probabilities of satisfied UEs.
	The baseline OLLA and the two eOLLA algorithms use the parameter $\Delta_{\text{up}}=0.5$ dB as assumed in many other performance studies \cite{sarret2015dynamicOLLA}. For the eOLLA Algorithms, we have simulated a series of instances to find the optimal value of $\Delta_{\text{down}}$ to maximize the number of satisfied users.
	We observed that the best case of the system capacity occurs when $\Delta_{\text{down}}$ is equal to $0.21$ and $0.044$ for Algorithms \ref{P1:Alg:Traditional_OLLA_per_CBG} and \ref{P1:Alg:eOLLA_ResidualBLER_2ndTx}, respectively. These values for $\Delta_{\text{down}}$ correspond to having a 1st-TX CBGER target of ~30\%, using (\ref{P1:Eq:CBGER_Convergence}) and the 2nd-TX TBER target of 15\% as per (\ref{P1:Eq:2ndTBER_Convergence}). This is considered a reasonable choice, given that the HARQ retransmissions only consist of a subset of the CBGs of the first transmission as observed also in Section \ref{P1:Sec3:Analytical_Assessment}.
	
	\subsection{Performance Results}\label{P1:Sec4:SubsecA:Performance Results}
\begin{figure}[t]
		\centerline{\includegraphics[width=\linewidth,trim={1.0cm 0.1cm 1.8cm 0.5cm},clip]{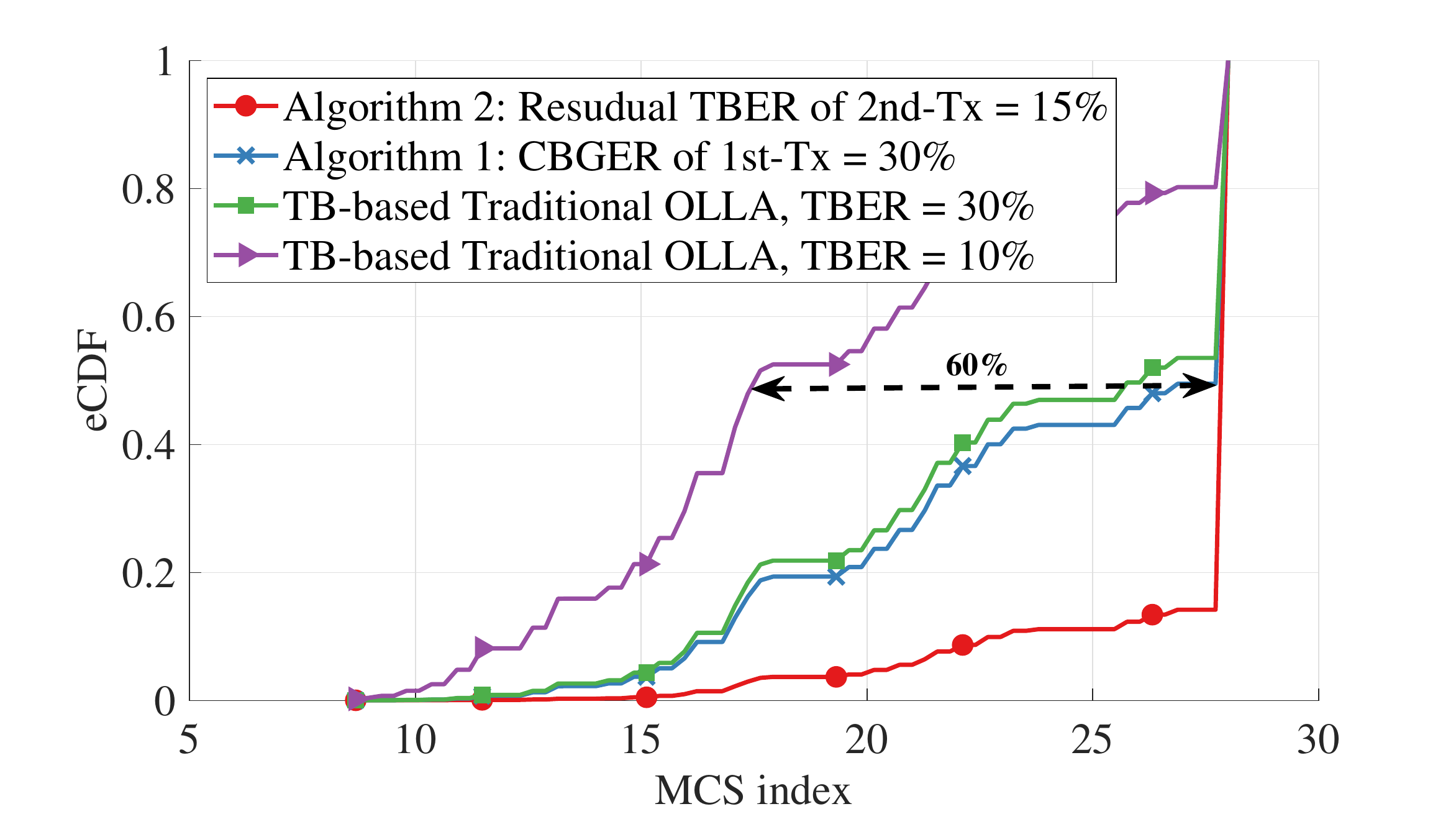}}
		\caption{eCDF of MCS index selection for all the OLLA variants with 5UEs/cell.}
		\label{P1:Fig:MCS_index}
	\end{figure}

	Fig. \ref{P1:Fig:MCS_index} shows the empirical cumulative distribution function (eCDF) of MCS index selection for different cases. 
	It can be seen that the eOLLA Algorithm \ref{P1:Alg:eOLLA_ResidualBLER_2ndTx} consistently selects higher MCS indices compared to the other schemes.
	As an example, eOLLA Algorithm \ref{P1:Alg:eOLLA_ResidualBLER_2ndTx} improves the MCS selection about \SI{60}{\percent} at 50\textsuperscript{th} percentile of eCDF compared to the TB-based traditional OLLA benchmark with \SI{10}{\percent} TBER target. This confirms the desired behavior where BS selects higher MCS indices for the CBG-based transmissions as this can be afforded, given that still only a fraction of the CBGs will fail in the first transmission,  resulting in more resource efficient transmissions.
	
\begin{figure}[t]
		\centerline{\includegraphics[width=\linewidth,trim={1.0cm 0.1cm 1.8cm 0.5cm},clip]{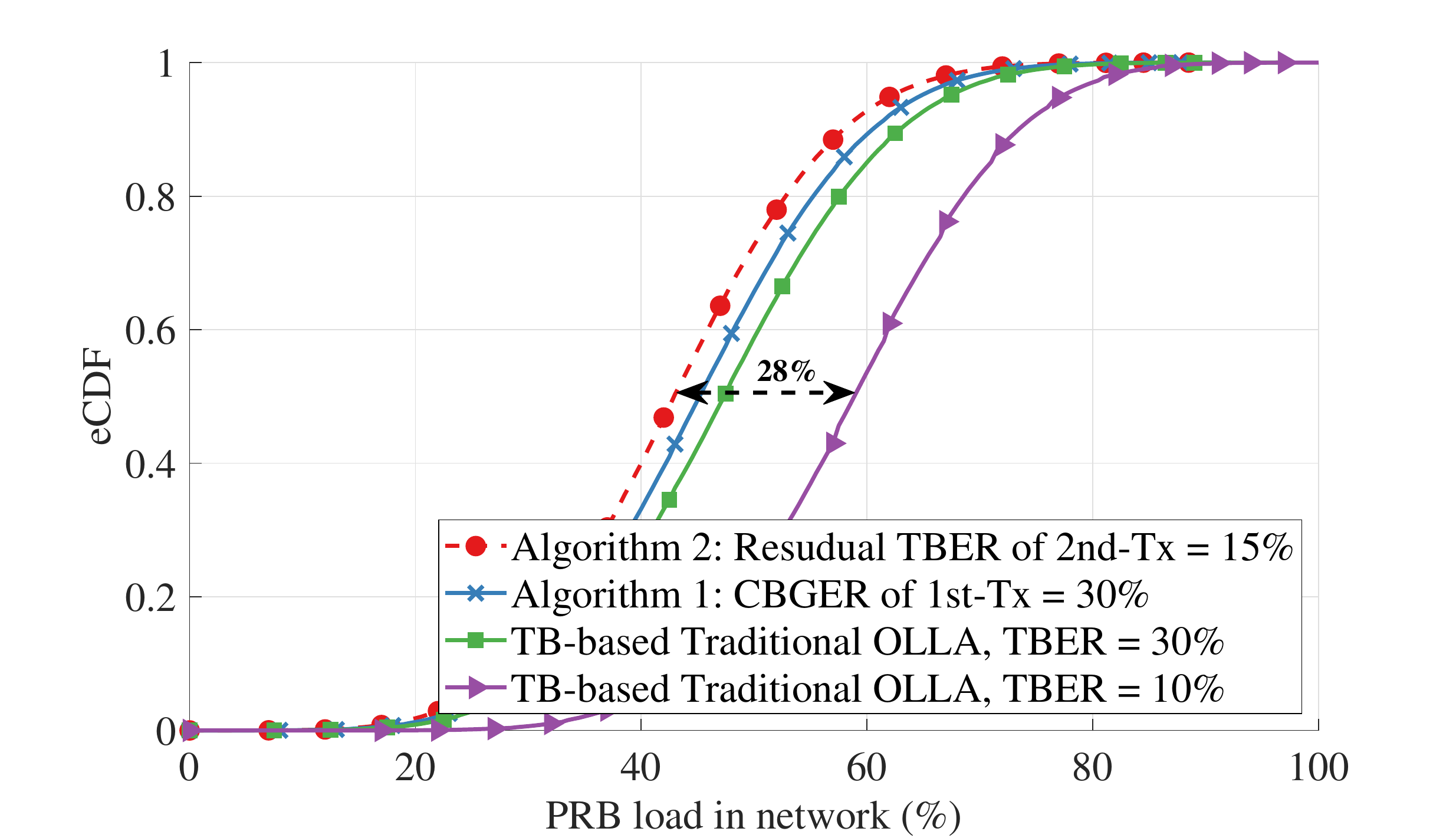}}
		\caption{eCDF of PRB utilization for the OLLA variants with 5 UEs/cell.}
		\label{P1:Fig:Average_RB_Load}
	\end{figure}

	Fig. \ref{P1:Fig:Average_RB_Load} illustrates the eCDF of PRB load in the network for different schemes when there are 5 UEs per cell. 
	It is observed that the proposed eOLLA algorithms outperform the baseline cases. The eOLLA Algorithm \ref{P1:Alg:eOLLA_ResidualBLER_2ndTx} saves about \SI{28}{\percent} radio resources at 50\textsuperscript{th} percentile of eCDF in comparison with TB-based traditional OLLA with \SI{10}{\percent} TBER target. The gain comes from fewer CBG retransmissions and higher MCS index selection that leads to an improved resource utilization. 
	Note that the result of system-level simulation results in Fig. \ref{P1:Fig:Average_RB_Load} match the theoretical RREG achieved by \eqref{P1:Eq:RSR} as shown in Fig. \ref{P1:Fig:ResourceEfficiencyGain_TBERandSINR} for the high SINR regime as is experienced in these system level simulations.
	
\begin{figure}[t]
		\centerline{\includegraphics[width=\linewidth,trim={1.0cm 0.0cm 1.8cm 0.5cm},clip]{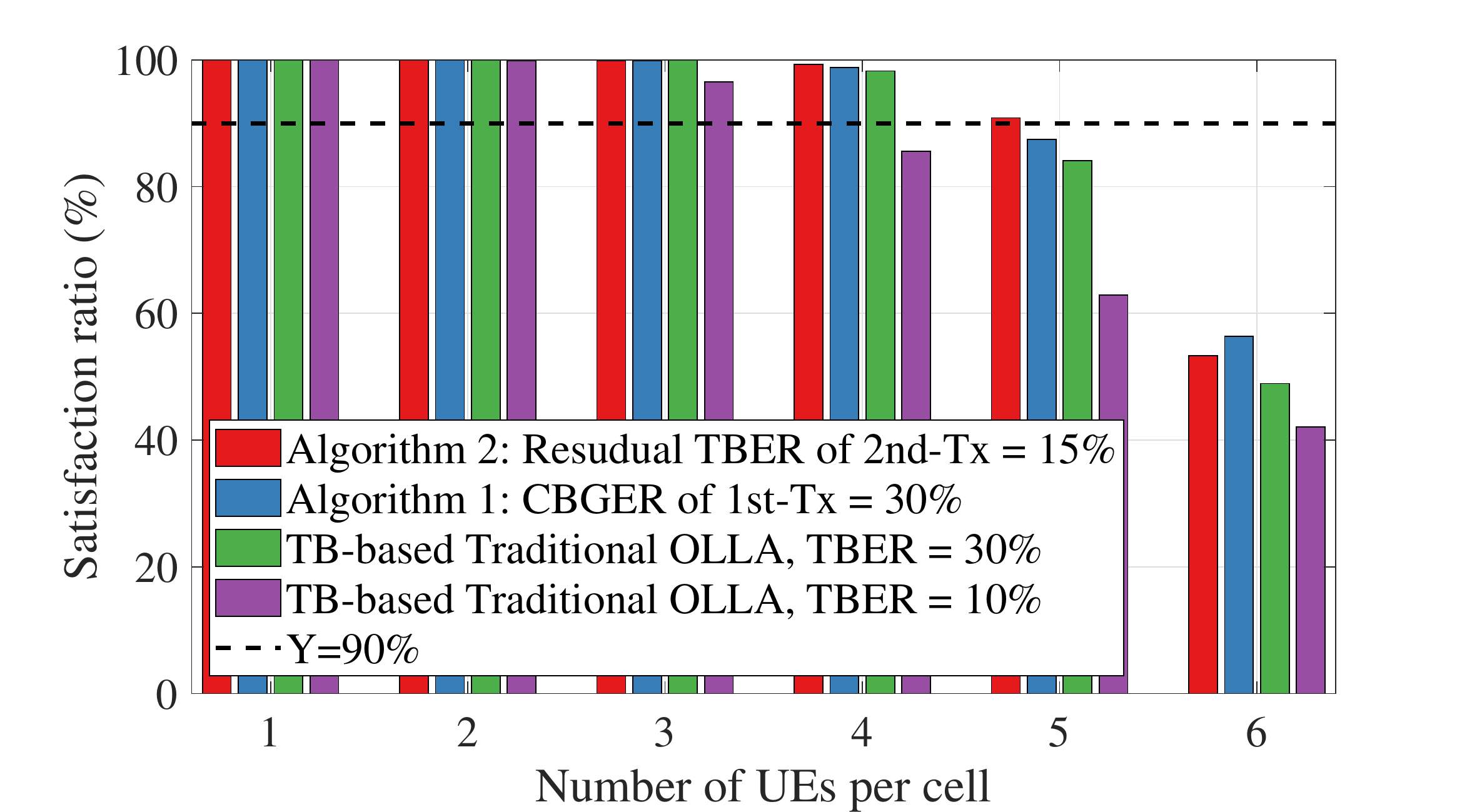}}
		\caption{System capacity: the supported number of UEs per cell when at least \SI{90}{\percent} of them are marked as satisfied UEs. A satisfied UE should successfully receive more than \SI{99}{\percent} of its own packets within 10 ms PDB \cite{3gpp.38.838}.}
		\label{P1:Fig:User_Happiness_ARVR_45Mbps}
	\end{figure}

	Fig. \ref{P1:Fig:User_Happiness_ARVR_45Mbps} depicts the system capacity of the network for different cases.
	Notice that both eOLLA algorithms outperform the TB-based traditional OLLA case.
	It can be seen that eOLLA Algorithm \ref{P1:Alg:eOLLA_ResidualBLER_2ndTx} supports two more XR UEs per cell as compared to the TB-based traditional OLLA with \SI{10}{\percent} TBER target. This is equivalent to a system capacity gain of \SI{67}{\percent}. This gain is materialized as a result of the improved resource utilization as reported in Fig. \ref{P1:Fig:Average_RB_Load}. It is furthermore observed that controlling the residual error of 2nd-TX is the most important, and hence eOLLA Algorithm \ref{P1:Alg:eOLLA_ResidualBLER_2ndTx} has better performance than eOLLA Algorithm \ref{P1:Alg:Traditional_OLLA_per_CBG}.
	
	\section{Conclusion}\label{P1:Sec6:Conclusion}
	In this paper, we proposed two novel low-complexity eOLLA algorithms tailored to CBG-based transmissions with multi-bit feedback for the XR downlink. The main takeaway of the study is that the eOLLA schemes are able to control the desired first and the second transmission error rate to improve resource efficiency of CBG-based transmissions, based on the multi-bit HARQ feedback. The design of the eOLLA schemes was guided by simple analytical framework. Dynamic system-level simulation results show \SI{33}{\percent} and \SI{67}{\percent} system capacity gains for the proposed Algorithm \ref{P1:Alg:Traditional_OLLA_per_CBG} and Algorithm \ref{P1:Alg:eOLLA_ResidualBLER_2ndTx}, respectively, in comparison with the TB-based traditional OLLA with \SI{10}{\percent} TBER target. The obtained simulation results match the simpler theoretical estimates fairly well. While both the eOLLA algorithms perform very well, results indicate that Algorithm \ref{P1:Alg:eOLLA_ResidualBLER_2ndTx} is the most promising solution as it controls the second transmission residual CBG error rate.

	\hyphenation{op-tical net-works semi-conduc-tor}
	\bibliographystyle{IEEEtran}
	\bibliography{IEEEabrv,Bibliography}
	
\end{document}

%% file: Commands.tex
\usepackage{cite}
\usepackage{amsmath,amssymb,amsfonts}
\usepackage{algorithmic}
\usepackage{graphicx}
\usepackage{textcomp}
\usepackage{xcolor}

\usepackage{lipsum,multicol}
\usepackage{enumitem}
\usepackage{algorithm}
\usepackage{algorithmic}
\usepackage{subfigure}
\usepackage[subfigure]{tocloft}
\usepackage[bottom]{footmisc}
\usepackage{multirow}
\usepackage{amsthm,amssymb,amsmath,bm}
\usepackage{rotating}
\usepackage{empheq}
\usepackage{booktabs}
\usepackage{notoccite}
\usepackage{amsmath,cases}
\usepackage{pdfpages}
\usepackage{etoolbox}
\usepackage{lastpage}
\usepackage{fancyhdr}
\usepackage{framed}
\usepackage{float}
\usepackage{subfigure}
\usepackage{comment}
\usepackage{tikz}
\usetikzlibrary{spy}
\usepackage{authblk}
\usepackage[normalem]{ulem}
\usepackage{soul}